# Perpendicularly magnetized YIG films with small Gilbert damping constant and anomalous spin transport properties

Qianbiao Liu[1,2], Kangkang Meng[1]*, Zedong Xu[3], Tao Zhu[4], Xiaoguang Xu[1], Jun Miao[1] & Yong Jiang[1]*

1. Beijing Advanced Innovation Center for Materials Genome Engineering, University of Science and Technology Beijing, Beijing 100083, China
2. Applied and Engineering physics, Cornell University, Ithaca, NY 14853, USA
3. Department of Physics, South University of Science and Technology of China, Shenzhen 518055, China
4. Institute of Physics, Chinese Academy of Sciences, Beijing 100190, China

Email: kkmeng@ustb.edu.cn; yjiang@ustb.edu.cn

**Abstract:** The $Y_3Fe_5O_{12}$ (YIG) films with perpendicular magnetic anisotropy (PMA) have recently attracted a great deal of attention for spintronics applications. Here, we report the induced PMA in the ultrathin YIG films grown on $(Gd_{2.6}Ca_{0.4})(Ga_{4.1}Mg_{0.25}Zr_{0.65})O_{12}$ (SGGG) substrates by epitaxial strain without preprocessing. Reciprocal space mapping shows that the films are lattice-matched to the substrates without strain relaxation. Through ferromagnetic resonance and polarized neutron reflectometry measurements, we find that these YIG films have ultra-low Gilbert damping constant ($α < 1×10^{-5}$) with a magnetic dead layer as thin as about 0.3 nm at the YIG/SGGG interfaces. Moreover, the transport behavior of the Pt/YIG/SGGG films reveals an enhancement of spin mixing conductance and a large non-monotonic magnetic field dependence of anomalous Hall effect as compared with the Pt/YIG/$Gd_3Ga_5O_{12}$ (GGG) films. The non-monotonic anomalous Hall signal is extracted in the temperature range from 150 to 350 K, which has been ascribed to the possible non-collinear magnetic order at the Pt/YIG interface induced by uniaxial strain.

The spin transport in ferrimagnetic insulator (FMI) based devices has received considerable interest due to its free of current-induced Joule heating and beneficial for low-power spintronics applications [1, 2]. Especially, the high-quality $Y_3Fe_5O_{12}$ (YIG) film as a widely studied FMI has low damping constant, low magnetostriction and small magnetocrystalline anisotropy, making it a key material for magnonics and spin caloritronics. Though the magnons can carry information over distances as long as millimeters in YIG film, there remains a challenge to control its magnetic anisotropy while maintaining the low damping constant [3], especially for the thin film with perpendicular magnetic anisotropy (PMA), which is very useful for spin polarizers, spin-torque oscillators, magneto-optical devices and magnon valves [4-7]. In addition, the spin-orbit torque (SOT) induced magnetization switching with low current densities has been realized in non-magnetic heavy metal (HM)/FMI heterostructures, paving the road towards ultralow-dissipation SOT devices based on FMIs [8-10]. Furthermore, previous theoretical studies have pointed that the current density will become much smaller if the domain structures were topologically protected (chiral) [11]. However, most FMI films favor in-plane easy axis dominated by shape anisotropy, and the investigation is eclipsed as compared with ferromagnetic materials which show abundant and interesting domain structures such as chiral domain walls and magnetic skyrmions *et al.* [12-17]. Recently, the interface-induced chiral domain walls have been observed in centrosymmetric oxides $Tm_3Fe_5O_{12}$ (TmIG) thin films, and the domain walls can be propelled by spin current from an adjacent platinum layer [18]. Similar with the TmIG films, the possible chiral magnetic structures are also expected in the YIG films with lower damping constant, which would further improve the chiral domain walls' motion speed.

Recently, several ways have been reported to attain the perpendicularly magnetized YIG films, one of which is utilizing the lattice distortion and magnetoelastic effect induced by epitaxial strain [19-22]. It is noted that the strain control can not only enable the field-free magnetization switching but also assist the stabilization of the non-collinear magnetic textures in a broad range of magnetic field and temperature. Therefore, abundant and interesting physical phenomena would emerge in epitaxial grown YIG films with PMA. However, either varying the buffer layer or doping would increase the Gilbert damping constant of YIG, which will affect the efficiency of the SOT induced magnetization switching [20, 21]. On the other hand, these preprocessing would lead to a more complicate magnetic structures and impede the further discussion of spin transport properties such as possible topological Hall effect (THE).

In this work, we realized the PMA of ultrathin YIG films deposited on SGGG substrates due to epitaxial strain. Through ferromagnetic resonance (FMR) and polarized neutron reflectometry (PNR) measurements, we have found that the YIG films had small Gilbert damping constant with a magnetic dead layer as thin as about 0.3 nm at the YIG/SGGG interfaces. Moreover, we have carried out the transport measurements of the Pt/YIG/SGGG films and observed a large non-monotonic magnetic field dependence of the anomalous Hall resistivity, which did not exist in the compared Pt/YIG/GGG films. The non-monotonic anomalous Hall signal was extracted in the temperature range from 150 to 350 K, and we ascribed it to the possible non-collinear magnetic order at the Pt/YIG interfaces induced by uniaxial strain.

**Results**

**Structural and magnetic characterization.** The epitaxial YIG films with varying thickness from 3 to 90 nm were grown on the [111]-oriented GGG substrates (lattice parameter $a$ = 1.237 nm) and SGGG substrates (lattice parameter $a$ = 1.248 nm) respectively by pulsed laser deposition technique (see methods). After the deposition, we have investigated the surface morphology of the two kinds of films using atomic force microscopy (AFM) as shown in Fig. 1 (a), and the two films have a similar and small surface roughness ~0.1 nm. Fig. 1 (b) shows the enlarged XRD $\omega$-$2\theta$ scan spectra of the YIG (40 nm) thin films grown on the two different substrates (more details are shown in the Supplementary Note 1), and they all show predominant (444) diffraction peaks without any other diffraction peaks, excluding impurity phases or other crystallographic orientations and indicating the single-phase nature. According to the (444) diffraction peak position and the reciprocal space map of the (642) reflection of a 40-nm-thick YIG film grown on SGGG as shown in Fig. 1(c), we have found that the lattice constant of SGGG (~1.248 nm) substrate was larger than the

YIG layer (~1.236 nm). We quantify this biaxial strain as $\xi = (a_{OP} - a_{IP})/a_{IP}$, where $a_{OP}$ and $a_{IP}$ represent the pseudo cubic lattice constant calculated from the out-of-plane lattice constant $d(4\ 4\ 4)_{OP}$ and in-plane lattice constant $d(1\ 1\ 0)_{IP}$, respectively, following the equation of $d = \dfrac{a}{\sqrt{h^2 + k^2 + l^2}}$, with $h$, $k$, and $l$ standing for the Miller indices of the crystal planes. It indicates that the SGGG substrate provides a tensile stress ($\xi \sim 0.84\%$) [21]. At the same time, the magnetic properties of the YIG films grown on the two different substrates were measured via VSM magnetometry at room temperature. According to the magnetic field (***H***) dependence of the magnetization (***M***) as shown in Fig. 1 (d), the magnetic anisotropy of the YIG film grown on SGGG substrate has been modulated by strain, while the two films have similar in-plane ***M-H*** curves.

To further investigate the quality of the YIG films grown on SGGG substrates and exclude the possibility of the strain induced large stoichiometry and lattice mismatch, compositional analyses were carried out using x-ray photoelectron spectroscopy (XPS) and PNR. As shown in Fig. 2 (a), the difference of binding energy between the $2p_{3/2}$ peak and the satellite peak is about 8.0 eV, and the Fe ions are determined to be in the $3^+$ valence state. It is found that there is no obvious difference for Fe elements in the YIG films grown on GGG and SGGG substrates. The Y $3d$ spectrums show a small energy shift as shown in Fig. 2 (b) and the binding energy shift may be related to the lattice strain and the variation of bond length [21]. Therefore, the stoichiometry of the YIG surface has not been dramatically modified with the strain control. Furthermore, we have performed the PNR measurement to probe the depth dependent structure and magnetic information of YIG films grown on SGGG substrates. The PNR signals and scattering length density (SLD) profiles for YIG (12.8 nm)/SGGG films by applying an in-plane magnetic field of 900 mT at room temperature are shown in Fig. 2 (c) and (d), respectively. In Fig. 2(c), $R_{++}$ and $R_{--}$ are the nonspin-flip reflectivities, where the spin polarizations are the same for the incoming and reflected neutrons. The inset of Fig. 2(c) shows the experimental and simulated spin-asymmetry (SA), defined as SA = $(R_{++} - R_{--})/(R_{++} + R_{--})$, as a function of scattering vector Q. A reasonable fitting was obtained with a three-layer model for the single YIG film, containing the interface layer, main YIG layer and surface layer. The nuclear SLD and magnetic SLD are directly proportional to the nuclear scattering potential and the magnetization, respectively. Then, the depth-resolved structural and magnetic SLD profiles delivered by fitting are shown in Fig. 2(d). The **Z**-axis represents the distance for the vertical direction of the film, where **Z** = 0 indicates the position at the YIG/SGGG interface. It is obvious that there is few Gd diffusion into the YIG film, and the dead layer (0.3 nm) is much thinner than the reported values (5-10 nm) between YIG (or TmIG) and substrates [23-25]. The net magnetization of YIG is 3.36 $\mu_B$ (~140 emu/cm$^3$), which is similar with that of bulk YIG [26]. The PNR results also showed that besides the YIG/SGGG interface region, there is also 1.51-nm-thick nonmagnetic surface layer, which may be $Y_2O_3$ and is likely to be extremely important in magnetic proximity effect [23].

**Dynamical characterization and spin transport properties.** To quantitatively determine the magnetic anisotropy and dynamic properties of the YIG films, the FMR spectra were measured at room temperature using an electron paramagnetic resonance spectrometer with rotating the films. Fig. 3(a) shows the geometric configuration of the angle resolved FMR measurements. We use the FMR absorption line shape to extract the resonance field ($H_{res}$) and peak-to-peak linewidth ($\Delta H_{pp}$) at different $\theta$ for the 40-nm-thick YIG films grown on GGG and SGGG substrates, respectively. The details for 3-nm-thick YIG film are shown in the Supplementary Note 2. According to the angle dependence of $H_{res}$ as shown in Fig. 3(b), one can find that as compared with the YIG films grown on GGG substrates, the minimum $H_{res}$ of the 40-nm-thick YIG film grown on SGGG substrate increases with varying $\theta$ from 0° to 90°. On the other hand, according to the frequency dependence of $H_{res}$ for the YIG (40 nm) films with applying **H** in the *XY* plane as shown in Fig. 3(c), in contrast to the YIG/GGG films, the $H_{res}$ in YIG/SGGG films could not be fitted by the in-plane magnetic anisotropy Kittel formula $f = (\gamma/2\pi)[H_{res}(H_{res} + 4\pi M_{eff})]^{1/2}$. All these results indicate that the easy axis of YIG (40 nm)/SGGG films lies out-of-plane. The angle dependent $\Delta H_{pp}$ for the two films are also compared as shown in Fig. 3(d), the 40-nm-thick YIG film grown on SGGG substrate has an optimal value of $\Delta H_{pp}$ as low as 0.4 mT at $\theta$=64°, and the corresponding FMR absorption line and Lorentz fitting curve are shown in Fig. 3(e). Generally, the $\Delta H_{pp}$ is expected to be minimum (maximum) along magnetic easy (hard) axis, which is basically coincident with the angle dependent $\Delta H_{pp}$ for the YIG films grown on GGG substrates. However, as shown in Fig. 3(d), the $\Delta H_{pp}$ for the YIG/SGGG films shows an anomalous variation. The lowest $\Delta H_{pp}$ at θ=64° could be ascribed to the high YIG film quality and ultrathin magnetic dead layer at the YIG/SGGG interface. It should be noted that, as compared with YIG/GGG films, the $\Delta H_{pp}$ is independent on the frequency from 5 GHz to 14 GHz as shown in Fig. 3(f). Then, we have calculated the Gilbert damping constant $α$ of the YIG (40 nm)/SGGG films by extracting the $\Delta H_{pp}$ at each frequency as shown in Fig. 3(f). The obtained $α$ is smaller than $1 \times 10^{-5}$, which is one order of magnitude lower than the report in Ref. [20] and would open new perspectives for the magnetization dynamics. According to the theoretical theme, the $\Delta H_{pp}$ consists of three parts: Gilbert damping, two magnons scattering relaxation process and inhomogeneities, in which both the Gilbert damping and the two magnons scattering relaxation process depend on frequency. Therefore, the large $\Delta H_{pp}$ in the YIG/SGGG films mainly stems from the inhomogeneities, which will be discussed next with the help of the transport measurements. All of the above results have proven that the ultrathin YIG films grown on SGGG substrates have not only evident PMA but also ultra-low Gilbert damping constant.

Furthermore, we have also investigated the spin transport properties for the high quality YIG films grown on SGGG substrates, which are basically sensitive to the magnetic details of YIG. The magnetoresistance (MR) has been proved as a powerful tool to effectively explore magnetic information originating from the interfaces [27]. The temperature dependent spin Hall magnetoresistance (SMR) of the Pt (5 nm)/YIG (3 nm) films grown on the two different substrates were measured using a small and non-perturbative current density (~ $10^6$ A/cm$^2$), and the sketches of the measurement is shown in Fig. 4 (a). The $\beta$ scan of the longitudinal MR, which is defined as MR=$\Delta\rho_{XX}/\rho_{XX(0)}$=[$\rho_{XX(\beta)}$ -$\rho_{XX(0)}$]/$\rho_{XX(0)}$ in the YZ plane for the two films under a 3 T field (enough to saturate the magnetization of YIG), shows $\cos^2\beta$ behaviors with varying temperature for the Pt/YIG/GGG and Pt/YIG/SGGG films as shown in Fig. 4 (b) and (c), respectively. The SMR of the Pt/YIG/SGGG films is larger than that of the Pt/YIG/GGG films with the same thickness of YIG at room temperature, indicating an enhanced spin mixing conductance ($G_{\uparrow\downarrow}$) in the Pt/YIG/SGGG films. Here, it should be noted that the spin transport properties for the Pt layers are expected to be the same because of the similar resistivity and films quality. Therefore, the SGGG substrate not only induces the PMA but also enhances $G_{\uparrow\downarrow}$ at the Pt/YIG interface. Then, we have also investigated the field dependent Hall resistivities in the Pt/YIG/SGGG films at the temperature range from 260 to 350 K as shown in Fig. 4(d). Though the conduction electrons cannot penetrate into the FMI layer, the possible anomalous Hall effect (AHE) at the HM/FMI interface is proposed to emerge, and the total Hall resistivity can usually be expressed as the sum of various contributions [28, 29]:

$$\rho_H = R_0 H + \rho_S + \rho_{S\text{-}A}, \qquad (1)$$

where $R_0$ is the normal Hall coefficient, $\rho_S$ the transverse manifestation of SMR, and $\rho_{S\text{-}A}$ the spin Hall anomalous Hall effect (SAHE) resistivity. Notably, the external field is applied out-of-plane, and $\rho_s$ (~$\Delta\rho_1 m_x m_y$) can be neglected [29]. Interestingly, the film grown on SGGG substrate shows a bump and dip feature during the hysteretic measurements in the temperature range from 260 to 350 K. In the following discussion, we term the part of extra anomalous signals as the anomalous SAHE resistivity ($\rho_{A\text{-}S\text{-}A}$). The $\rho_{A\text{-}S\text{-}A}$ signals clearly coexist with the large background of normal Hall effect. Notably, the broken (space) inversion symmetry with strong spin-orbit coupling (SOC) will induce the Dzyaloshinskii-Moriya interaction (DMI). If the DMI could be compared with the Heisenberg exchange interaction and the magnetic anisotropy that were controlled by strain, it could stabilize non-collinear magnetic textures such as skyrmions, producing a fictitious magnetic field and the THE. The $\rho_{A\text{-}S\text{-}A}$ signals indicate that a chiral spin texture may exist, which is similar with B20-type compounds $Mn_3Si$ and $Mn_3Ge$ [30,31]. To more clearly demonstrate the origin of the anomalous signals, we have subtracted the normal Hall term, and the temperature dependence of ($\rho_{S\text{-}A} + \rho_{A\text{-}S\text{-}A}$) has been shown in Fig. 4 (e). Then, we can further discern the peak and hump structures in the temperature range from 260 to 350 K. The SAHE contribution $\rho_{S\text{-}A}$ can be expressed as $\rho_{S-A} = \Delta\rho_2 m_Z$ [32, 33], where $\Delta\rho_2$ is the coefficient depending on the imaginary part of $G_{\uparrow\downarrow}$, and $m_z$ is the unit vector of the magnetization orientation along the **Z** direction. The extracted Hall resistivity $\rho_{A\text{-}S\text{-}A}$ has been shown in Fig. 4 (f), and the temperature dependence of the largest $\rho_{A\text{-}S\text{-}A}$ ($\rho_{A-S-A}^{\text{Max}}$) in all the films have been shown in Fig. 4 (g). Finite values of $\rho_{A-S-A}^{\text{Max}}$ exist in the temperature range from 150 to 350 K, which is much different from that in B20-type bulk chiral magnets which are subjected to low temperature and large magnetic field [34]. The large non-monotonic magnetic field dependence of anomalous Hall resistivity could not stem from the Weyl points, and the more detailed discussion was shown in the Supplementary Note 3.

To further discuss the origin of the anomalous transport signals, we have investigated the small field dependence of the Hall resistances for Pt (5 nm)/YIG (40 nm)/SGGG films as shown in Fig. 5(a). The out-of-plane hysteresis loop of Pt/YIG/SGGG is not central symmetry, which indicates the existence of an internal field leading to opposite velocities of up to down and down to up domain walls in the presence of current along the +$X$ direction. The large field dependences of the Hall resistances are shown in Fig. 5(b), which could not be described by Equation (1). There are large variations for the Hall signals when the external magnetic field is lower than the saturation field ($B_s$) of YIG film (~50 mT at 300 K and ~150 mT at 50 K). More interestingly, we have firstly applied a large out-of-plane external magnetic field of +0.8 T (-0.8 T) above $B_s$ to saturate the out-of-plane magnetization component $M_Z > 0$ ($M_Z < 0$), then decreased the field to zero, finally the Hall resistances were measured in the small field range (± 400 Oe), from which we could find that the shape was reversed as shown in Fig. 5(c). Here, we infer that the magnetic structures at the Pt/YIG interface grown on SGGG substrate could not be a simple linear magnetic order. Theoretically, an additional chirality-driven Hall effect might be present in the ferromagnetic regime due to spin canting [35-38]. It has been found that the strain from an insulating substrate could produce a tetragonal distortion, which would drive an orbital selection, modifying the electronic properties and the magnetic ordering of manganites. For $A_{1-x}B_xMnO_3$ perovskites, a compressive strain makes the ferromagnetic configuration relatively more stable than the antiferromagnetic state [39]. On the other hand, the strain would induce the spin canting [40]. A variety of experiments and theories have reported that the ion substitute, defect and magnetoelastic interaction would cant the magnetization of YIG [41-43]. Therefore, if we could modify the magnetic order by epitaxial strain, the non-collinear magnetic structure is expected to emerge in the YIG film. For YIG crystalline structure, the two Fe sites are located on the octahedrally coordinated 16(a) site and the tetrahedrally coordinated 24(d) site, aligning antiparallel with each other [44]. According to the XRD and RSM results, the tensile strain due to SGGG substrate would result in the distortion angle of the facets of the YIG unit cell smaller than 90° [45]. Therefore, the magnetizations of Fe at two sublattices should be discussed separately rather than as a whole. Then, the anomalous signals of Pt/YIG/SGGG films could be ascribed to the emergence of four different $Fe^{3+}$ magnetic orientations in strained Pt/YIG films, which are shown in Fig. 5(d). For better to understand our results, we assume that, in analogy with $\rho_S$, the $\rho_{A-S-A}$ is larger than $\rho_{A-S}$ and scales linearly with $m_y m_z$ and $m_x m_z$. With applying a large external field $H$ along $Z$ axis, the uncompensated magnetic moment at the tetrahedrally coordinated 24(d) is along with the external fields $H$ direction for $|H| > B_s$, and the magnetic moment tends to be along $A$ (-$A$) axis when the external fields is swept from 0.8 T (-0.8 T) to 0 T. Then, if the Hall resistance was measured at small out-of-plane field, the uncompensated magnetic moment would switch from $A$(-$A$) axis to $B$(-$B$) axis. In this case, the $\rho_{A-S-A}$ that scales with $\Delta\rho_3(m_y m_z + m_x m_z)$ would change the sign because the $m_z$ is switched from the **Z** axis to -**Z** axis as shown in Fig. 5(c). However, there is still some problem that needs to be further clarified. There are no anomalous signals in Pt/YIG/GGG films that could be ascribed to the weak strength of $\varDelta\rho_3$ or the strong magnetic anisotropy. It is still valued for further discussion of the origin of $\varDelta\rho_3$ that whether it could stem from the skrymions *et al.*, but until now we have not observed any chiral domain structures in Pt/YIG/SGGG films through the Lorentz transmission electron microscopy. Therefore, we hope that future work would involve more detailed magnetic microscopy imaging and microstructure analysis, which can further elucidate the real microscopic origin of the large non-monotonic magnetic field dependence of anomalous Hall resistivity.

**Conclusion**

In conclusion, the YIG film with PMA could be realized using both epitaxial strain and growth-induced anisotropies. These YIG films grown on SGGG substrates had low Gilbert damping constants ($<1\times10^{-5}$) with a magnetic dead layer as thin as about 0.3 nm at the YIG/SGGG interface. Moreover, we observed a large non-monotonic magnetic field dependence of anomalous Hall resistivity in Pt/YIG/SGGG films, which did not exist in Pt/YIG/GGG films. The non-monotonic anomalous portion of the Hall signal was extracted in the temperature range from 150 to 350 K and we ascribed it to the possible non-collinear magnetic order at the Pt/YIG interface induced by uniaxial strain. The present work not only demonstrate that the strain control can effectively tune the electromagnetic properties of FMI but also open up the exploration of non-collinear spin texture for fundamental physics and magnetic storage technologies based on FMI.

**Methods**

**Sample preparation.** The epitaxial YIG films with varying thickness from 3 to 90 nm were grown on the [111]-oriented GGG substrates (lattice parameter $a$ =1.237 nm) and SGGG substrates (lattice parameter $a$ =1.248 nm) respectively by pulsed laser deposition technique. The growth temperature was $T_S$ =780 °C and the oxygen pressure was varied from 10 to 50 Pa. Then, the films were annealed at 780°C for 30 min at the oxygen pressure of 200 Pa. The Pt (5nm) layer was deposited on the top of YIG films at room temperature by magnetron sputtering. After the deposition, the electron beam lithography and Ar ion milling were used to pattern Hall bars, and a lift-off process was used to form contact electrodes. The size of all the Hall bars is 20 μm×120 μm.

**Structural and magnetic characterization.** The surface morphology was measured by AFM (Bruke Dimension Icon). Magnetization measurements were carried out using a Physical Property Measurement System (PPMS) VSM. A detailed investigation of the magnetic information of YIG was investigated by PNR at the Spallation Neutron Source of China.

**Ferromagnetic resonance measurements.** The measurement setup is depicted in Fig. 3(a). For FMR measurements, the DC magnetic field was modulated with an AC field. The transmitted signal was detected by a lock-in amplifier. We observed the FMR spectrum of the sample by sweeping the external magnetic field. The data obtained were then fitted to a sum of symmetric and antisymmetric Lorentzian functions to extract the linewidth.

**Spin transport measurements.** The measurements were carried out using PPMS DynaCool.

**Acknowledgments**

The authors thanks Prof. L. Q. Yan and Y. Sun for the technical assistant in ferromagnetic resonance measurement. This work was partially supported by the National Science Foundation of China (Grant Nos. 51971027, 51927802, 51971023, 51731003, 51671019, 51602022, 61674013, 51602025), and the Fundamental Research Funds for the Central Universities (FRF-TP-19-001A3).

**Figure Captions**

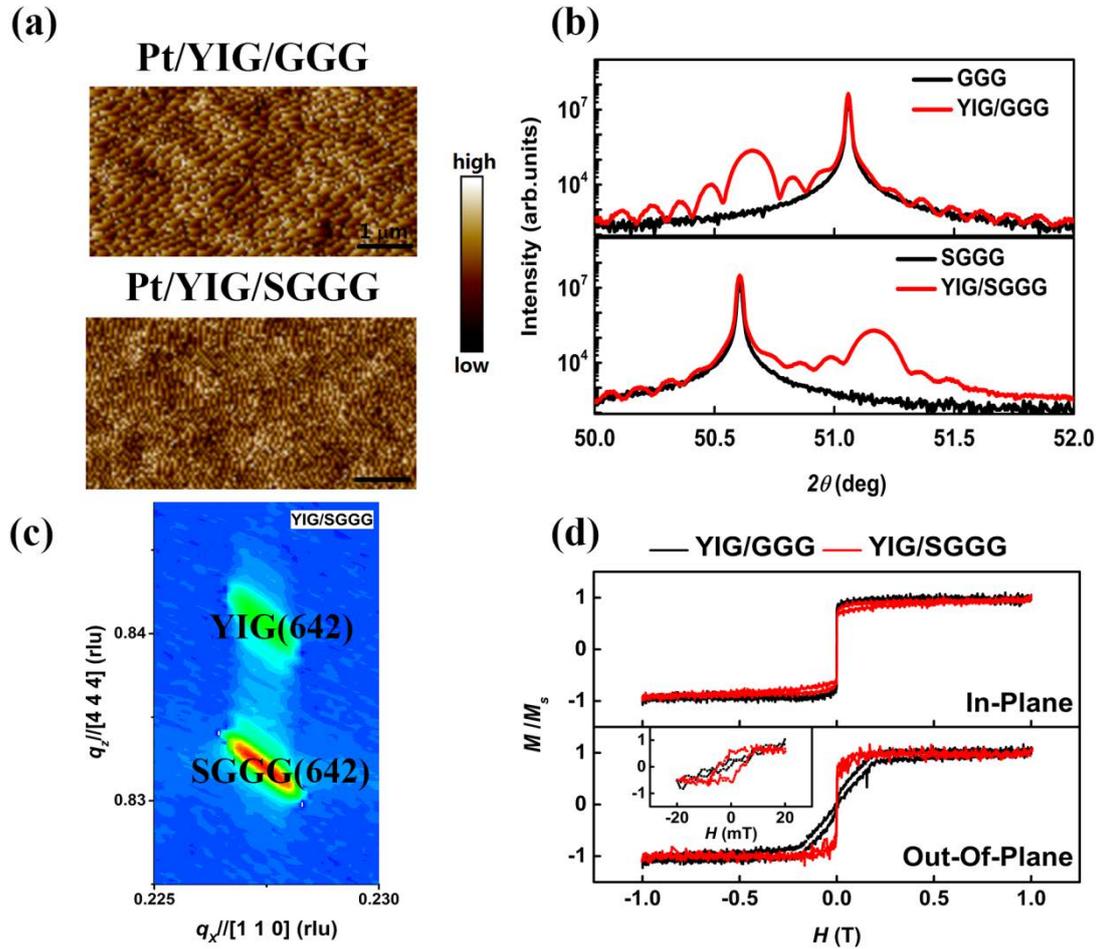

**Fig. 1 Structural and magnetic properties of YIG films.** (a) AFM images of the YIG films grown on the two substrates (scale bar, 1 μm). (b) XRD $\omega$-$2\theta$ scans of the two different YIG films grown on the two substrates. (c) High-resolution XRD reciprocal space map of the YIG film deposited on the SGGG substrate. (d) Field dependence of the normalized magnetization of the YIG films grown on the two different substrates.

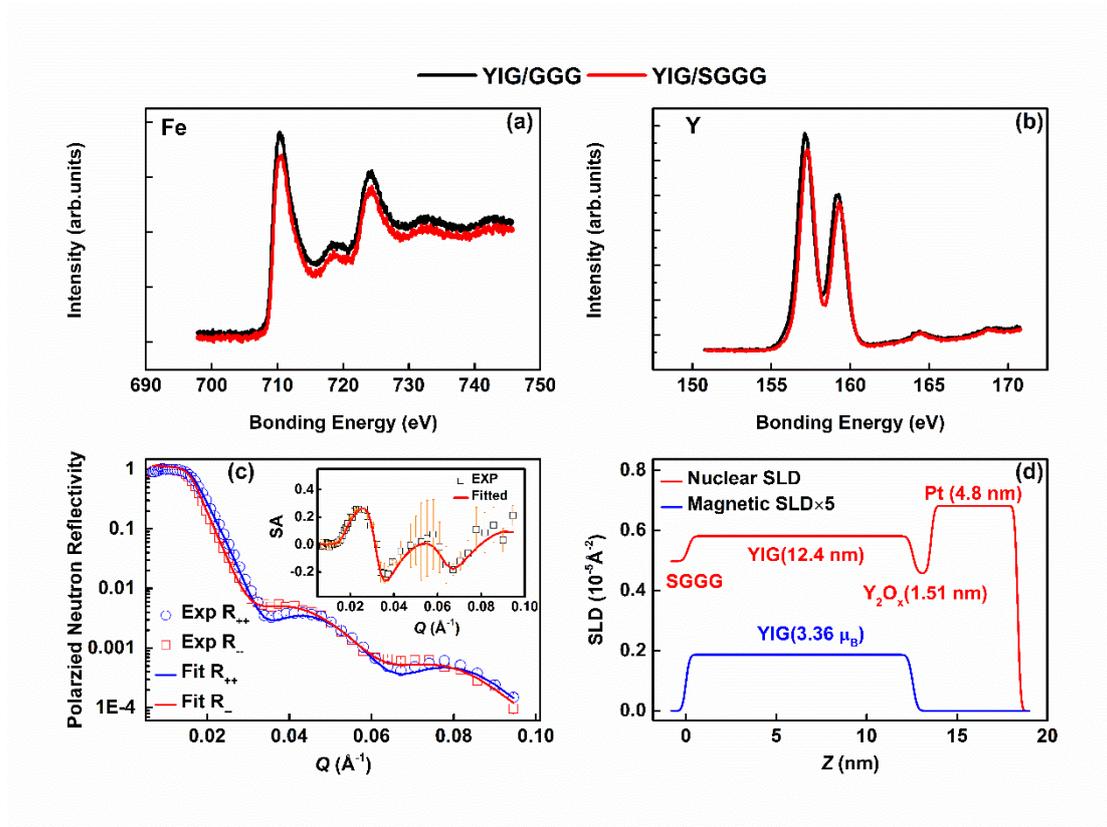

**Fig. 2 Structural and magnetic properties of YIG films.** Room temperature XPS spectra of (a) Fe *2p* and (b) Y *3d* for YIG films grown on the two substrates. (c) PNR signals (with a 900 mT in-plane field) for the spin-polarized $R_{++}$ and $R_{--}$ channels. Inset: The experimental and simulated SA as a function of scattering vector Q. (d) SLD profiles of the YIG/SGGG films. The nuclear SLD and magnetic SLD is directly proportional to the nuclear scattering potential and the magnetization, respectively.

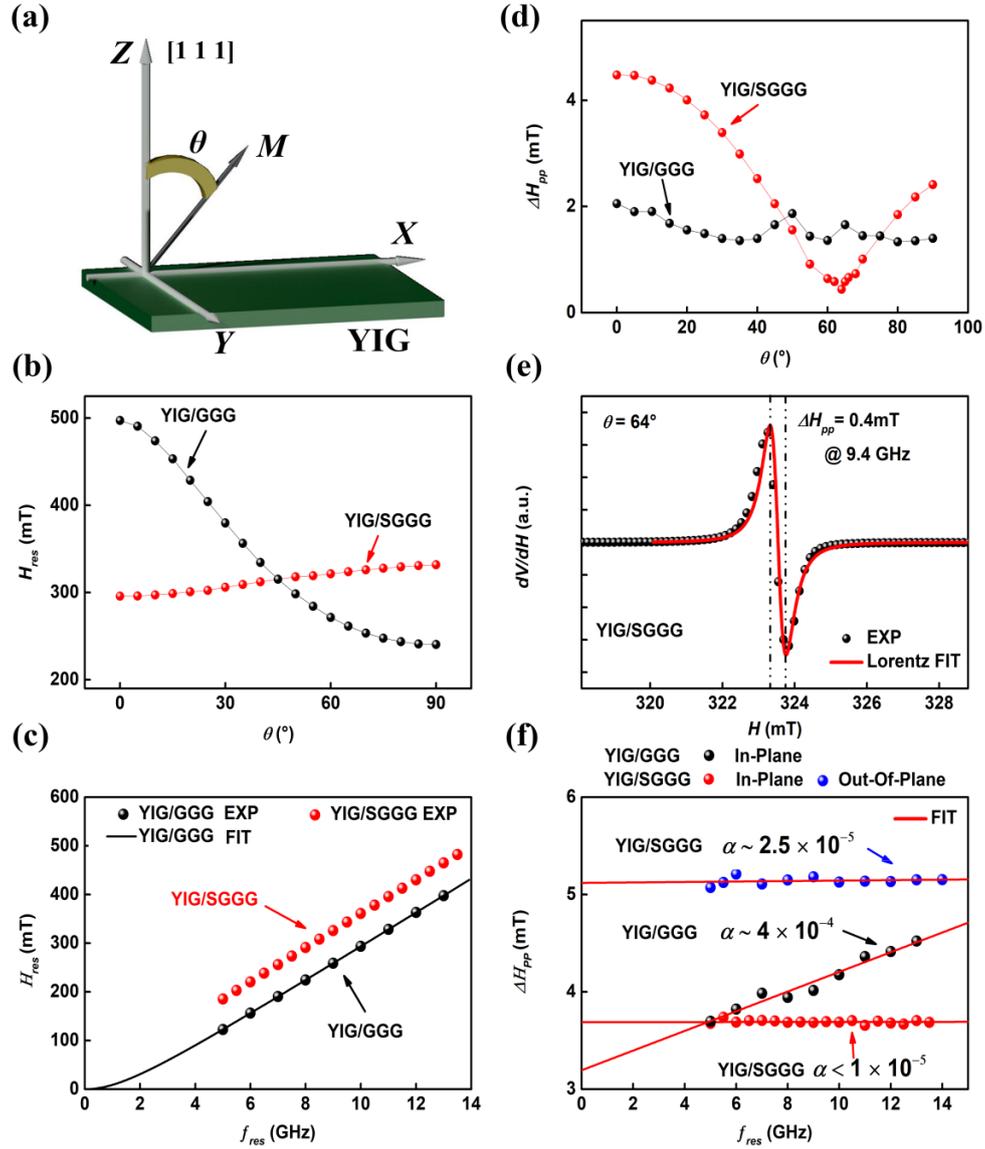

**Fig. 3 Dynamical properties of YIG films.** (a) The geometric configuration of the angle dependent FMR measurement. (b) The angle dependence of the $H_{res}$ for the YIG films on GGG and SGGG substrates. (c) The frequency dependence of the $H_{res}$ for YIG films grown on GGG and SGGG substrates. (d) The angle dependence of $\Delta H_{pp}$ for the YIG films on GGG and SGGG substrates. (e) FMR spectrum of the 40-nm-thick YIG film grown on SGGG substrate with 9.46 GHz at $\theta=64°$. (f) The frequency dependence of $\Delta H_{pp}$ for the 40-nm-thick YIG films grown on GGG and SGGG substrates.

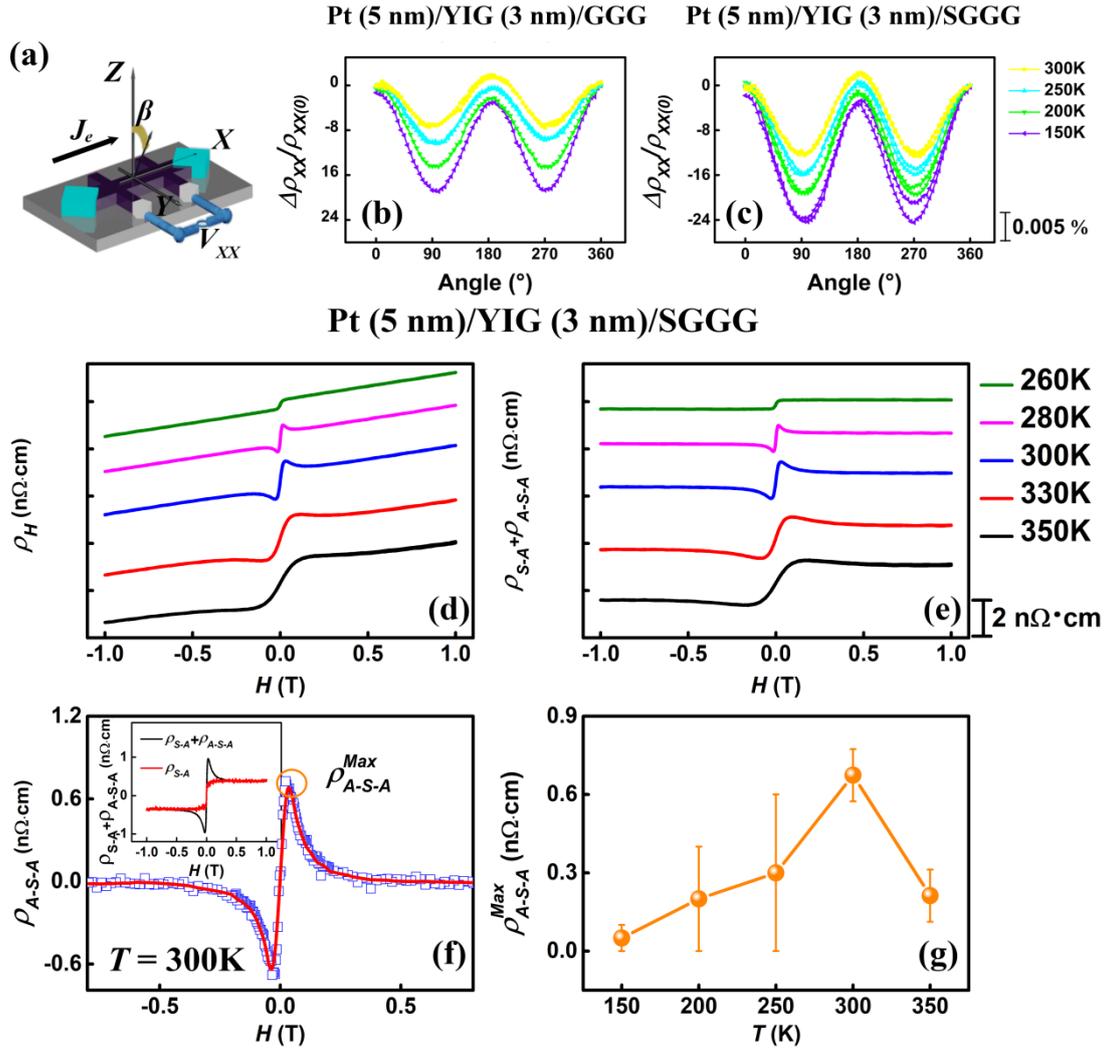

**Fig. 4 Spin transport properties of Pt/YIG (3nm) films.** (a) The definition of the angle, the axes and the measurement configurations. (b) and (c) Longitudinal MR at different temperatures in Pt/YIG/GGG and Pt/YIG/SGGG films respectively (The applied magnetic field is 3 T). (d) Total Hall resistivities *vs* $H$ for Pt/YIG/SGGG films in the temperature range from 260 to 300 K. (e) ($\rho_{S-A}+\rho_{A-S-A}$) *vs* $H$ for two films in the temperature range from 260 to 300 K. (f) $\rho_{A-S-A}$ *vs* $H$ for Pt/YIG/SGGG films at 300K. Inset: $\rho_{S-A}$ and $\rho_{S-A} + \rho_{A-S-A}$ *vs* $H$ for Pt/YIG/SGGG films at 300K. (g) Temperature dependence of the $\rho_{A-S-A}^{Max}$.

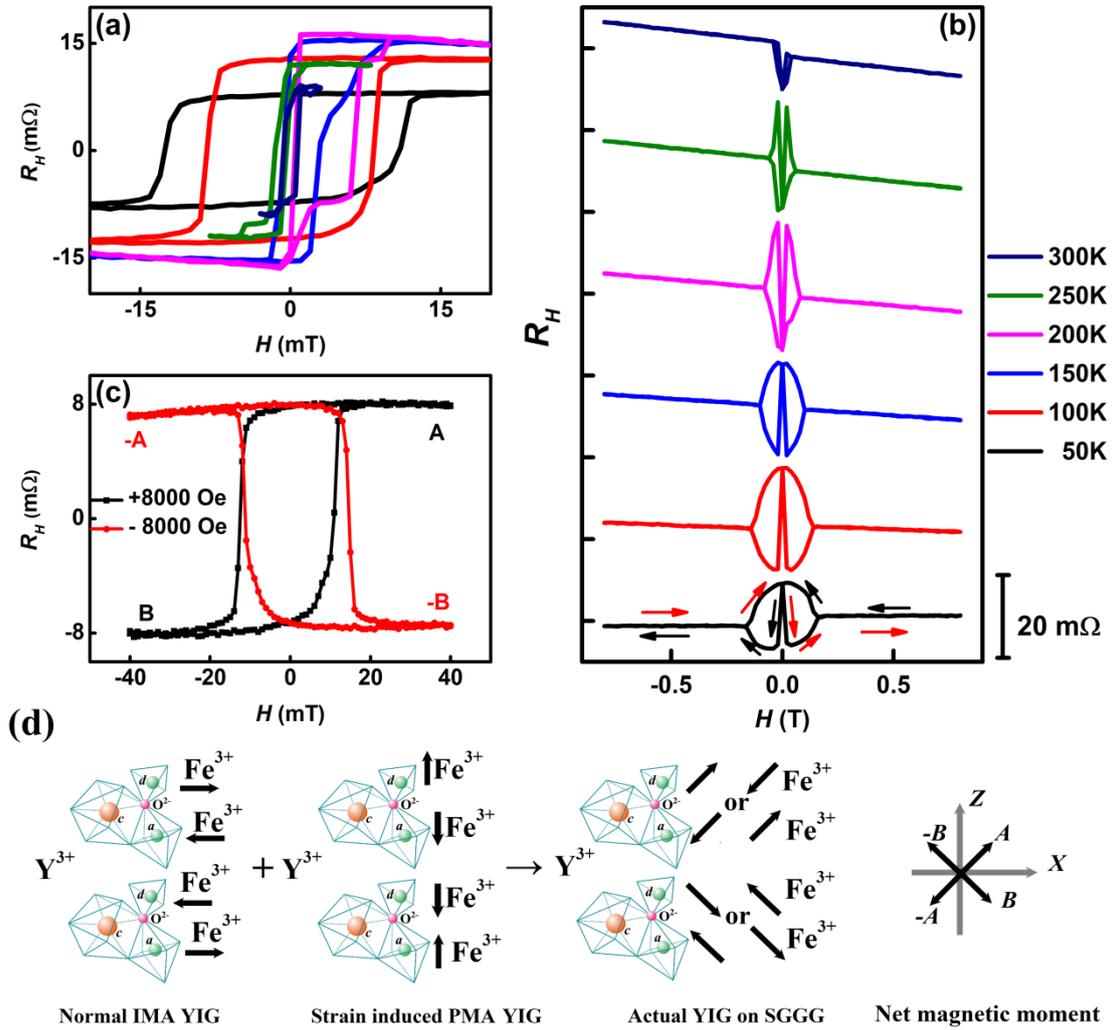

Figure 5 **Spin transport properties of Pt/YIG (40 nm) films.** (a) and (b) The Hall resistances *vs* **H** for the Pt/YIG/SGGG films in the temperature range from 50 to 300 K in small and large magnetic field range, respectively. (c) The Hall resistances *vs* **H** at small magnetic field range after sweeping a large out-of-plane magnetic field +0.8 T (black line) and -0.8 T (red line) to zero. (d) An illustration of the orientations of the magnetizations Fe (*a*) and Fe (*d*) in YIG films with the normal in-plane magnetic anisotropy (IMA), the ideal strain induced PMA and the actual magnetic anisotropy grown on SGGG in our work.